\documentclass[aps,prl,twocolumn,superscriptaddress,groupedaddress]{revtex4}
\usepackage{subfig}
\usepackage{graphicx}  
\usepackage{dcolumn}   
\usepackage{bm}        
\usepackage{amssymb}   
\usepackage{slashed}
\usepackage{graphicx}				
\usepackage{amsmath}
\usepackage{mathtools}
\usepackage{tikz,pgf}
\usepackage{comment}
\usepackage{asymptote}
\usetikzlibrary{arrows,backgrounds}
\usetikzlibrary{fit,scopes,calc,matrix,positioning,decorations.pathmorphing}
\usepackage[all]{xy}
\usepackage{yfonts}

\newcommand{\ket}[1]{\ensuremath{\left|#1\right\rangle}}

\DeclareFontFamily{OMS}{oasy}{\skewchar\font48 }
\DeclareFontShape{OMS}{oasy}{m}{n}{%
         <-5.5> oasy5     <5.5-6.5> oasy6
      <6.5-7.5> oasy7     <7.5-8.5> oasy8
      <8.5-9.5> oasy9     <9.5->  oasy10
      }{}
\DeclareFontShape{OMS}{oasy}{b}{n}{%
       <-6> oabsy5
      <6-8> oabsy7
      <8->  oabsy10
      }{}
\DeclareSymbolFont{oasy}{OMS}{oasy}{m}{n}
\SetSymbolFont{oasy}{bold}{OMS}{oasy}{b}{n}

\DeclareMathSymbol{\smallleftarrow}     {\mathrel}{oasy}{"20}
\DeclareMathSymbol{\smallrightarrow}    {\mathrel}{oasy}{"21}
\DeclareMathSymbol{\smallleftrightarrow}{\mathrel}{oasy}{"24}

\begin{document}
\title{Emergence of non-abelian gauge theories and the ER-EPR duality }
\author{Andrei T. Patrascu}
\address{ELI-NP, Horia Hulubei National Institute for R\&D in Physics and Nuclear Engineering, 30 Reactorului St, Bucharest-Magurele, 077125, Romania}
\begin{abstract}
Making use of the emergence of an entangling gate from the Mayer Vietoris maps generating a torus and employing the correspondence between quantum error correction codes and gauge symmetries I show that non-abelian gauge theories are related to the existence of entanglement. This observation is based on the fact that string compactification on a torus provides non-trivial fluxes (gaugings) when the twists are chosen such that the obtained torus matches the maps arising in the Mayer-Vietoris theorem which result in entangling gates. This observation adds new indications towards the fact that low energy non-abelian gauge theories emerge from entanglement related phenomena.
\end{abstract}
\maketitle
Quantum field theory faced many problems in its early days. One of the first was the difficulty associated with understanding interactions. While a free theory was easily formulated and solved, interactions were not easily incorporated. The way we understand interactions nowadays is through the so called gauge field theories. Requiring gauge invariance means requiring the invariance of the theory's Lagrangian to local unmeasurable changes in the phase of our quantum fields. In some sense, gauge invariance is equivalent to the independence of the theory to local arbitrary choices of frames for the quantum fields. Making such invariance manifest requires certain corrections to the Lagrangian that lead to interactions. Hence, basically quantum interactions are a natural result of gauge invariance. 
But what is the origin of this gauge invariance or gauge symmetry? It certainly is a quantum effect, as it relies on the invariance of the phase of our fields' phases, but it can also be recovered in classical electrodynamics in a somewhat different form. While of course in classical electrodynamics we also have gauge invariance, understanding it in its full complexity with non-abelian symmetry generators requires the existence of quantum fields and their phases. It appears that we can give a quantum information meaning to it starting even at the most fundamental string theoretical level, by means of entanglement and the ER-EPR duality. Indeed, it seems like our gauge fields and the interactions they mediate are a "condensed" form of quantum entanglement at the fundamental high energy level. 
While in quantum field theory, particularly in quantum electrodynamics, the entanglement can be associated either to $s$ channel processes where the virtual photon carries equal overlaps of helicities of the final state particles, or to the superposition between $t$ and $u$ channels [19], it appears clear that the condition of requiring the possibility of generating maximal entanglement has different results whether it is applied to QED or to the weak or strong interaction. Within a UV incomplete, non-renormalisable quantum theory of gravity, we can imagine that the graviton will also carry entanglement, however the structure of this entanglement as well as possible constraints on the ability of producing maximum entanglement will have different forms, due to the tensorial nature of the graviton. In any case it seems that there is an underlying connection between the possibility of generating maximal entanglement and the fundamental gauge symmetries of high energy physics. 
The generation of entangled photons by means of graviton coupling in perturbative non-renormalisable quantum gravity has been studied to some extend in previous articles [20]. However, string theory has the capacity of showing different, more extensive entanglement structures than what can be envisioned via perturbative low energy quantum gravity. Also, fixing the gauge in string theory usually involves a BRST approach. Hence, one can expect a connection between the BRST cohomology and the entanglement measured in a string theory, by means, for example, of the replica trick. This article relies on a more heuristic approach, in which the topological connection of ER=EPR is used to qualitatively derive the connection between non-abelian gauge symmetry and entanglement, leaving for a future article to show it explicitly in terms of the connection between BRST cohomology and entanglement. However, this article is in a sense more general, as it presents an overarching homological algebraic technique that derives both the generic connection between entanglement and topology and, via an argument from string compactification, the connection between quantum error correction algorithms and gauge symmetry. 

The observation that toroidal topology leads to entangling effects has been shown in [1] following the observation by [2]. It is also a fundamental result of ref. [18].  A typical error correction code consists of several entangling gates protecting the information stored in qubits from various types of deterioration. It has been shown in [3] that there exists a dual relationship between quantum error correction codes and the redundancy offered by gauge invariance. The duality employed in this context was holographic in nature. In this article I will show that a dual relation between non-abelian gauge theories and entanglement can be found by looking at the ER-EPR correspondence instead. This relation implies however string compactification and brings the high energy behaviour into focus.  New deep connections between quantum information theory and gauge field theories have been the subject of extended research in [5-10]. The subject is still developing considering the fact that a better understanding of entanglement in the context of quantum field theory is a relatively recent development [11]. 
In standard quantum field theory, particularly in the various theories of the standard model, QED, electroweak, and strong interaction, it has been shown that the origin of gauge interactions can be linked to entanglement and that the requirement of maximal entanglement leads to gauge-like behaviour in these theories [19]. 

The study of the dictionary lying behind the holographic duality has been under scrutiny since the introduction of the AdS/CFT duality [12]. This has led to a better understanding of the black hole information problem [13], the introduction of precursor operators[14], and the derivation of the entanglement entropy formula in a holographic context by Ryu and Takayanagi [15].  The study of entanglement in the context of the holographic principle has led to additional quantum corrections to this formula, as discovered in [16] and to a new interpretation of those results from a topological perspective in [17]. On the other side, the research on the newly formulated ER-EPR correspondence still waits for a proper understanding of the underlying physics. While there have been several attempts to implement an ER-EPR dictionary following the model of the AdS/CFT dictionary, the development is far less advanced.  
In this article I propose a better understanding of entanglement in a string theoretical context by following the Scherk-Schwarz compactification from the perspective of the ER-EPR duality as derived by means of the Mayer-Vietoris theorem [1]. The ER-EPR duality states that a wormhole or a topologically non-trivial spacetime automatically entails entanglement. Considering that the Scherk Schwarz compactification on a torus implies the fact that several space directions become compact and take the shape of a torus with non-trivial cycles, it appears that the ER-EPR duality may be applicable. The surprising result is that the non-abelian gauge structure of the effective theory must emerge from the intrinsic entanglement entailed by the non-trivial topological structure of the compact spacetime directions. It is particularly intriguing, albeit it is left for a future work, to understand similar entanglement-gauge connections in the context of Calabi-Yau compactifications. Entanglement as needed for quantum error correction codes offers precisely the type of redundancy required for understanding the emergence of non-abelian gauge theories by means of non-vanishing fluxes in torus compactification.  The main technique used here is the Scherk-Schwarz compactification and the way in which geometric fluxes emerge when it is employed.  The underlying theory we start with is supergravity albeit, I will also describe possible non-geometric expansions in the context of double field theory. Given the NS-NS sector of supergravity with the $D$-dimensional metric $g_{ij}$, the 2-form field $b_{ij}$ and a dilaton $\phi$ depending on $D$ spacetime coordinates $x^{i}$, we may dimensionally reduce it to $d=D-n$ dimensions resulting in some effective theory. In order to do this, we split the coordinates, selecting which will correspond to compact space dimensions and which will correspond to the large effective theory dimensions. The notation used corresponds to [4] and is 
\begin{equation}
x^{i}= (x^{\mu},y^{m})
\end{equation}
 with $m= 1,...,n$ being the compact space dimensions called internal, and $\mu= 1,...,d$ are the spacetime directions in the effective theory, also called external.  The fields of the original higher-dimensional theory belong to the representations of the symmetry groups of that theory. The process of compactification decomposes the original fields into the representations of the lower dimensional symmetry groups

\begin{equation}
g_{ij}=\left( {\begin{array}{cc}
	g_{\mu\nu}+g_{pq}A^{p}_{\;\;\mu}A^{q}_{\;\;\nu} & A^{p}_{\;\;\mu}g_{pn}\\
	g_{mp}A^{p}_{\;\;\nu} & g_{mn}
	\end{array}}\right)
\end{equation}

and for the 2-form field
\begin{widetext}
\begin{equation}
b_{ij}=\left( {\begin{array}{cc}
b_{\mu\nu}-\frac{1}{2}(A^{p}_{\;\;\nu}V_{p\nu}-A^{p}_{\;\;\nu}V_{p\mu})+A^{p}_{\;\;\mu}A^{q}_{\;\;\nu}b_{pq} & V_{n\mu}-b_{np}A^{p}_{\;\;\mu}\\
	-V_{m\nu}+b_{mp}A^{p}_{\;\;\nu} & b_{mn} \end{array}}\right)
\end{equation}
\end{widetext}
The fields therefore will appear to be formed out of external, internal, and mixed components. The gauge parameters also do split 

\begin{equation}
\lambda^{i}= (\epsilon^{\mu},\Lambda^{m}),\tilde{\lambda}_{i}= (\epsilon_{\mu},\Lambda_{m})
\end{equation}

A global shift symmetry of the type $b\rightarrow b+v$ can simply be gauged by allowing it to depend on the internal coordinates $v\rightarrow v(y)$, and hence introducing the twists $u(y)$ and $v(y)$. After finalisation of the compactification procedure the dependence on the internal coordinates will disappear, yet the information on the twist will remain in the form of structure-like constants in the effective theory. To see the split between the internal (compact) and external directions explicitly let me introduce the dependence on the internal (compact) directions by means of the functions $u^{\;\;m}_{a}(y)$ and $v_{mn}(y)$ resulting in a re-writing of the fields of the form 
\begin{widetext}
\begin{equation}
\begin{array}{ll}
g_{\mu\nu}= \hat{g}_{\mu\nu}(x), & b_{\mu\nu}=\hat{b}_{\mu\nu}(x)\\
A^{m}_{\mu}=u^{m}_{a}(y)\hat{A}^{a}_{\mu}(x), & V_{m\mu}=u^{a}_{\;\;m}(y)\hat{V}_{a\mu}(x)\\
g_{mn}=u^{a}_{\;\;m}(y) u^{b}_{\;\;n}(y)\hat{g}_{ab}(x), & b_{mn}=u^{a}_{m}(y)u^{b}_{\;\;n}(y)\hat{b}_{ab
}(x) + v_{mn}(y)\\
\lambda^{i} = (\hat{\epsilon}^{\mu}(x),u^{\;\;m}_{a}(y)\hat{\Lambda}^{a}(x)), & \hat{\lambda}_{i}= (\hat{\epsilon}_{\mu}(x), u^{\;\;a}_{m}(y)\hat{\Lambda}_{a}(x))
\end{array}
\end{equation}
\end{widetext}

and $\phi=\hat{\phi}(x)$ for the dilaton.  The twist matrices are taken to be constant in the external directions in order to preserve Lorentz invariance in the effective action. The hatted fields on the other side depend only on the external coordinates and correspond to dynamical degrees of freedom in the effective action.  The twists depending on the internal degrees of freedom on the other hand will play a particularly important role in the context of the ER-EPR duality as derived by means of the Mayer-Vietoris theorem. Indeed, Mayer-Vietoris theorem shows that when a torus is being constructed out of objects of simpler cohomology, the maps patching those pieces together amount to the entangler gate. In the case of bipartite entanglement the emerging entangler gate is particularly simple and will generate the standard Hadamard-CNOT gate.  It is well known that the quantum error correction codes, from the simple 3-qubit error correction code, to the 9-qubit quantum error correction code developed by Shor rely on the successive application of entangler gates relating various possible quantum states. While the 3-qubit code cannot simultaneously correct both for bit and phase flips, the 9-qubit code can and is therefore considered to be truly quantum. The quantum error correction mechanism generates a series of redundancies by coupling ancillas to the original code. Those are generally constructed such that they detect whether an error has occurred and correct it. The 3-qubit code is capable of encoding one single logical qubit into three physical qubits such that it automatically can correct for a single bit flip. The basis states for the logical qubit are defined as
\begin{equation}
\begin{array}{cc}
\ket{0}_{L}=\ket{000}, & \ket{1}_{L}=\ket{111}\\
\end{array}
\end{equation}

Given an arbitrary qubit state
$\ket{\psi}=\alpha \ket{0}+\beta\ket{1}$
we may represent it in this basis as
\begin{equation}
\begin{array}{c}
\ket{\psi}=\alpha\ket{0}+\beta\ket{1} \rightarrow \\
\rightarrow \alpha\ket{0}_{L}+\beta\ket{1}_{L}=\alpha\ket{000}+\beta\ket{111}=\ket{\psi}_{L}
\end{array}
\end{equation}

A single logical qubit can be encoded by initialising two ancilla qubits and two CNOT gates. Two additional ancilla qubits are used to extract the syndrome information regarding possible errors from the data block without being able to identify the exact state of any qubit.  The correction procedure introduces two ancilla qubits again and performs a sequence of CNOT gates checking the parity of the three qubit data block. It is obvious that in order to identify and correct the errors in the logical qubit state, there appears to be a requirement for a special type of redundancy related to entanglement. The distribution of the logical information over the entangled qubits is found in the construction of a torus from spaces of smaller topological complexity. Compactification on a genus-one torus is the simple case to be considered here, and hence the Mayer-Vietoris theorem will be used on such a torus. However, Calabi-Yau compactification may reveal new insights on the connection between the emergence of gauge fields and quantum error correction codes. Indeed, the study of Calabi-Yau compactification from the perspective of the Mayer-Vietoris theorem with the focus on the type of emerging entanglement will be the subject of a future article. For our torus, the Mayer Vietoris theorem states that the cohomology of a complicated space may be described in terms of simpler sub-spaces defined on the original space provided one takes into account the way the intersection of these subspaces map both in the simple subspaces and in the large resulting space itself. The Mayer-Vietoris theorem for the most general case can simply be stated in terms of a long exact sequence
\begin{widetext}
\begin{equation}
\begin{array}{c}
...\rightarrow H_{n+1}(X)\xrightarrow{\partial_{*}} H_{n}(A\cap B)\xrightarrow{(i_{*},j_{*})}H_{n}(A)\oplus Hn(B)\xrightarrow{k_{*}-l_{*}}H_{n}(X)\xrightarrow{\partial_{*}}H_{n-1}(A\cap B)\rightarrow\\
...\rightarrow H_{0}(A)\oplus H_{0}(B)\xrightarrow{k_{*}-l_{*}}H_{0}(X)\rightarrow 0\\
\end{array}
\end{equation}
\end{widetext}

The original space $X$ is split into the topologically simpler subspaces $A$ and $B$ with their topology determined by the respective groups $H_{n}(A)$ and $H_{n}(B)$.  The cohomology of the intersection of the simpler subspaces is $H_{n}(A\cap B)$ and finally, the direct sum of the cohomology of the two subspaces partially encodes the whole torus. The definition of the maps arising in this long exact sequence is the same as in [1] and the long exact sequence, when particularised to a finite dimensional torus will only contain the subsequences corresponding to the relevant internal dimensions. What is however important here is the Mayer-Vietoris map
\begin{equation}
H_{n}(A\cap B)\xrightarrow{(i_{*},j_{*})}H_{n}(A)\oplus H_{n}(B)
\end{equation}

where $(i_{*},j_{*})$ is induced in homology by the inclusions $i:A\cap B\hookrightarrow A$ and $j:A\cap B\hookrightarrow B$. This map is not an isomorphism, neither when acting on the original space, nor when acting on homologies [1]. It has been shown that this map generates a Hadamard gate when a global twist in the torus is introduced. It has also been shown in [1] that the next map, $(k_{*}-l_{*})$ introduces an additional twist on the torus compensating for the one introduced by the first inclusion map. While the first inclusion map plays the role of the Hadamard gate, the second one plays the role of CNOT, resulting in an untwisted torus with an entanglement structure on it.  It is clear that twist operators play a particularly important role in making the intrinsic entanglement of the torus manifest. Indeed, it has been shown in [1] that this step represents the formal interpretation of the ER-EPR duality. The fact that a certain form of redundancy found in quantum error correction codes may be equated with gauge invariance has been observed in [2].  In the theory studied here, namely supergravity, we have the gauge transformations given by the Lie derivatives
\begin{equation}
L_{\lambda}V^{\mu}=\lambda^{j}\partial_{j}V^{i}-V^{j}\partial_{j}\lambda^{i}
\end{equation}

as acting on the vector $V^{\mu}$. In addition to this we have the gauge transformations of the 2-form

\begin{equation}
 b_{ij}\rightarrow b_{ij}+\partial_{i}\tilde{\lambda}_{j}-\partial_{j}\tilde{\lambda}_{i}
 \end{equation}
 
 In the effective theory, with
  $V^{i}= (\hat{v}^{\mu}(x),u^{\;\;m}_{a}(y)\hat{v}^{a}(x))$ 
  the gauge transformation becomes 
  \begin{equation}
  L_{\lambda}V^{\mu}=\hat{\epsilon}^{\nu}\partial_{\nu} \hat{v}^{\mu}-\hat{v}^{\nu}\partial_{\nu}\hat{\epsilon}^{\mu}=\hat{L}_{\hat{\epsilon}}\hat{v}^{\mu} 
 \end{equation}
 In therms of the twist coefficients we have 
 
 \begin{equation}
 L_{\lambda}V^{m}=u^{m}_{a}\hat{L}_{\hat{\lambda}}\hat{V}^{a}
 \end{equation}
 
and we can gauge the resulting transformation
\begin{equation}
\hat{L}_{\hat{\lambda}}\hat{V}^{a}=L_{\hat{\lambda}}\hat{V}^{a}+f^{\;\;\;\;a}_{bc}\hat{\Lambda}^{b}\hat{v}^{c}
\end{equation}

The contribution to this gauged transformation comes from the combination of twist matrices given by 

\begin{equation} 
f^{\;\;\;c}_{ab}=u^{\;\;m}_{a}\partial_{m}u^{\;\;n}_{b}u^{c}_{\;\;n}-u^{m}_{b}\partial_{m} u^{n}_{a}u^{c}_{\;\;n}
\end{equation}

These objects are known as geometric fluxes. One imposes the constraint that they are constant in the external coordinates.  They are also called metric fluxes as they correspond to the background fluxes of the metric. On the other side, $u_{a}^{\;\;m}(y)$ corresponds to the internal coordinate dependence of the metric. In a shorthand notation and following [4] we can write them as 

\begin{equation}
\begin{array}{c}
f_{abc}= 3(\partial_{[a}v_{bc]}+f^{d}_{[ab}v_{c]d})\\
f^{\;\;\;c}_{ab}=u^{\;\;m}_{a}\partial_{m}u^{\;\;n}_{b}u^{c}_{\;\;n}-u^{\;\;m}_{b}\partial_{m}u^{\;\;n}_{a}u^{c}_{\;\;n}\\
\end{array}
\end{equation}

and the rest, namely $f^{\;\;bc}_{a}=0$, $f^{abc}=0$. Compactification on a torus with vanishing background of the 2-form corresponds to $u^{a}_{\;\;m}=\delta^{a}_{m}$ and $v_{ab}= 0$ [4] which leads to $f_{abc}= 0$ and hence an ungauged theory.  With the twists turned on and depending on the internal coordinate we can make a choice such that $f_{abc}\neq 0$ which leads to a non-abelian gauge theory and possibly a cosmological constant term. But the twist matrices are precisely those matrices defined on the torus and depending on the internal coordinates. Their choice is not determined when the torus is already formed, but they must be con-strained whenever we intend to form a torus from surfaces of lower complexity. In fact, these twist matrices are determined from the choices required in the Mayer-Vietoris theorem. Moreover, their non-trivial nature is precisely equivalent to the demand that the two relevant maps in the Mayer-Vietoris sequence lead to entanglement. Heuristically, the simple surfaces that need to be patched together require an initial twist that would generate what is interpreted in quantum information theory as the Hadamard gate generating the state
\begin{equation}
\ket{0} \rightarrow \frac{1}{\sqrt{2}}(\ket{0_{A}}+\ket{1_{B}})
\end{equation}

and finally another twist that would compensate for the first one that would be associated with the CNOT operation on the lower qubit leading, via direct product, to the entangled state. Indeed, a similar procedure can be found in the Scherk-Schwarz compactification prescription where the original dependence on the original coordinates is explicitly eliminated at the end of the procedure in order to preserve Lorentz invariance on the external directions, yet, the twist remains and has an impact on the internal coordinates, playing the role of structure-like constants in the effective theory. Indeed, this strongly suggests that the non-abelian nature of effective field theories emerges from topological entanglement at the level of high energy geometry and topology. As a conclusion, I noticed in this paper that the non-abelian gauge symmetry structure is related to high energy supergravity entanglement and that the Mayer-Vietoris derivation of the ER-EPR duality allows us to better understand the source of non-abelian gauge invariance in a geometric flux environment.  This work is at its first stages.Understanding the structure of entanglement via similar methods in the context of Calabi-Yau manifolds would result in an improved understanding of the string vacua problem as well as a possible understanding on how the cosmological constant could emerge from string theoretical arguments. The same analysis can be considered in non-geometrical backgrounds by replacing our supergravity with a double field theory. In this context T-duality will be explicitly considered and will result in various alterations of the way in which Mayer-Vietoris sequences are applied.  One has to consider the non-Riemannian nature of the geometry involved as well as the fact that the patching between subspaces must be done by means of maps that incorporate T-duality. This will be the subject of a future research. The main result of this paper, namely that non-abelian gauge theories have their origin in toroidal (bipartite) entanglement for the compactification geometry in the Scherk-Schwarz context is new and may provide novel research directions in the future.

\end{document}